\documentclass[%
 reprint,
 showpacs,preprintnumbers,
 amsmath,amssymb,
 aps,
 prb,
]{revtex4-1}

\usepackage{graphicx}
\usepackage{dcolumn}
\usepackage{bm}

\begin{document}

\preprint{APS/123-QED}

\title{
First Principle Computation of Random Pinning Glass Transition, Glass Cooperative Length-Scales and Numerical Comparisons}
\thanks{This project has received funding
from the European Union’s Horizon 2020 research and innovation programme under the Marie Sklodowska-Curie
grant agreement No. 654971, the ERC grant CRIPHERASY (no. 247328) and from the MINECO (Spain) (contract No. FIS2012-35719-C02). Also we
  wish to thank S. Franz, G. Biroli and G. Tarjus for helpful discussions
  in the first stage of this work.}%

\author{Chiara Cammarota}
 \email{chiara.cammarota@roma1.infn.it}
\affiliation{
Mathematics Department, King's College London, Strand WC2R 2LS, United Kingdom
}
\affiliation{%
 Physics Department of ``La Sapienza'', University of Rome,
 P.le Aldo Moro 2, 00185 Rome, Italy
}%

\author{Beatriz Seoane}
\email{beatriz.seoane.bartolome@lpt.ens.fr}
\affiliation{Laboratoire de Physique Th\'eorique, \'Ecole Normale Sup\'erieure, UMR 8549 CNRS, 24 Rue Lhomond, 75005 Paris, France}
\affiliation{%
 Physics Department of ``La Sapienza'', University of Rome,
 P.le Aldo Moro 2, 00185 Rome, Italy
}%
\date{\today}

\begin{abstract}
As a guideline for experimental tests of the ideal glass
transition (Random Pinning Glass Transition, RPGT) that shall be induced in a
system by randomly pinning particles, we performed first-principle
computations within the Hypernetted chain approximation and numerical simulations
of a Hard Sphere model of glass-former. We obtain confirmation of the expected
enhancement of glassy behaviour under the procedure of random pinning, which consists in
freezing a fraction $c$ of randomly chosen particles in the positions they
have in an equilibrium configuration.
We present the analytical phase diagram as a function of $c$ and of the packing
fraction $\phi$, showing a line of RPGT ending in a critical point.
We also obtain first microscopic results on cooperative length-scales characterizing medium-range
amorphous order in Hard Spere glasses and indirect
quantitative information on a key thermodynamic quantity defined in proximity of ideal 
glass transitions, the amorphous surface tension. 
Finally, we present numerical results of pair correlation functions able to differentiate the liquid and the glass phases, as predicted by the analytic computations.
\end{abstract}

\pacs{Valid PACS appear here}
\maketitle

A common feature of liquids deep below the melting point (supercooled)
is the remarkably steep increase of relaxation time, until the system falls
out of equilibrium at a conventional temperature $T_\mathrm{G}$.
Since decades, it is present in the literature the claim of the possible
presence of a phase transition\cite{cavagn09,berbir11}, the ideal glass
transition (IGT), underlying the dynamical arrest and located at a lower
temperature $T_K$. Despite the first formulation of a consistent 
phenomenological thermodynamic picture\cite{kitiwo89} this intuition remained 
at length debated for 
the lack of other indicators of the imminent thermodynamic singularity. 
In this context, other theoretical perspectives based on dynamic or
topological approaches, sometimes excluding the presence of any transition,
have been proposed as alternative explanations of the sluggish
dynamics\cite{chagar10,takinv05}.\\
In recent times, new important results have been obtained
in the development of the thermodynamic scenario. We particularly refer
on the one hand, to the first definition and measure 
of a new kind of cooperative 
length-scale\cite{boubir04,bibcgv08} detecting the 
spatial extent of amorphous order\cite{kurlev11}, 
called point-to-set $l_{PS}$, and on the other hand,
to the development of a field theory description of
the IGT in terms of a suitable large deviation
function\cite{frapar98,berthi13,parisi14} and the introduction of
perturbative\cite{frparr2011,rizzot13} and 
non-perturbative\cite{cabitt11,bicatt13} fluctuations in this description.
In view of a full fledged theory of glass-formers and quantitative
predictions of their properties, these advancements have
been corroborated by the formulation of a microscopic 
theory\cite{mezpar99,cafrpa98,cafrpa99,parzam10} 
inspired to classical first-principle computations techniques in
liquids\cite{morhir61,hanmcd90}. \\
Due to particularly severe critical properties of the 
IGT, {\it i.e.} an exponential growth of relaxation time and a power law increase of the cooperative length,
the revealing of its properties still remains too difficult to be achieved, leaving fundamental doubts 
on the whole theoretical picture and on the existence 
of the IGT itself.\\
With the aim of giving an answer to these fundamental questions one of us
recently proposed\cite{cb12PNAS,cambir13} a procedure to induce in real systems a glass
transition of more easy access than the IGT.
The idea is as follows: freeze the position of a fraction $c$ of particles of
an equilibrated configuration and study the thermodynamics of the remaining
free particles.
This procedure should allow the observation of a remarkable growth of the
relaxation time in the free-particles equilibrium dynamics and the
eventual reaching of a glass transition, called Random Pinning Glass
Transition (RPGT), as soon as the concentration of frozen particles reaches
a critical value $c_K$. This result is valid in a
full range of moderate and deep supercooling leading to the formation of a
line of RPGT, $c_K(T)$, ending in the IGT at $T_K$ and in a new
glass critical point at high temperature.
This suggests that the way for a test of the theory of
IGT is open and calls for numerical\cite{kobber13,karpar13,kalepr12,chchta12,kobcos14} 
and experimental studies of the phase diagram of glass-formers with a fraction of frozen particles. 
On the other side, a
microscopic first-principles theory of the RPGT is needed to give
more detailed predictions on its physics and provide quantitative
information useful in the planning stage of experimental and numerical tests.

In this rapid communication we report the results of a first-principle
computation of the RPGT scenario following the method for a quantitative 
approach to IGT proposed in Ref.~\onlinecite{mezpar99}: an extension to glasses
of the Hypernetted Chain (HNC) approximation of classical theory of simple
liquids. 
Moreover, we report numerical simulation results 
to be compared with analytic predictions.
As glass-former model 
we chose an Hard Sphere system in $3$ dimensions, for which a microscopic 
thermodynamic theory has already been developed\cite{parzam10} in the
unconstrained case. In HS models, particles interact through the usual HS
potential $\psi(r)=0$ for $r \geq D_{ij}$ and $\infty$ for $r<D_{ij}$, where
$D_{ij}$ is the sum of the radii $R_{i,j}$ of particles $i$ and $j$. 
HNC computations refer to a monodisperse HSs system. 
The control parameter ruling the approach to IGT is the density, $\rho=N/V$ 
for a system with $N$ particles in a volume $V$, or the packing 
fraction $\phi=4\pi R^3\rho/3$: the fraction of volume occupied by the particles. 
To prevent crystallization, numerical simulations have instead be realised for a $50:50$ binary mixture of $N=250$ type A and B spheres of radius $R_B=1.4R_A$ and packing fraction $\phi=2\pi (R_A^3+R_B^3)\rho/3$.\\ As anticipated in the introduction, the emergence of medium 
range fluctuations of amorphous order in proximity of the IGT 
is a long standing open issue.
Despite many 
efforts\cite{boubir04,monsem06,framon07,cabitt11}, quantitative 
information on spatial correlation of glass order remained out of 
reach of the HNC and of other first principles 
approaches to realistic glass-formers.
The present computation, combining together the HNC analytic approach with the pinning particle
procedure, at last overcomes this limitation obtaining 
first quantitative predictions on diverging 
length-scales of the unconstrained system and 
even some indirect results on the amorhpous surface tension, a thermodynamic 
quantity expected to play a major role in the vicinity of the IGT.\\

\begin{figure}
\hspace{-0.3cm}
\includegraphics[height=0.35\textwidth, angle=0]{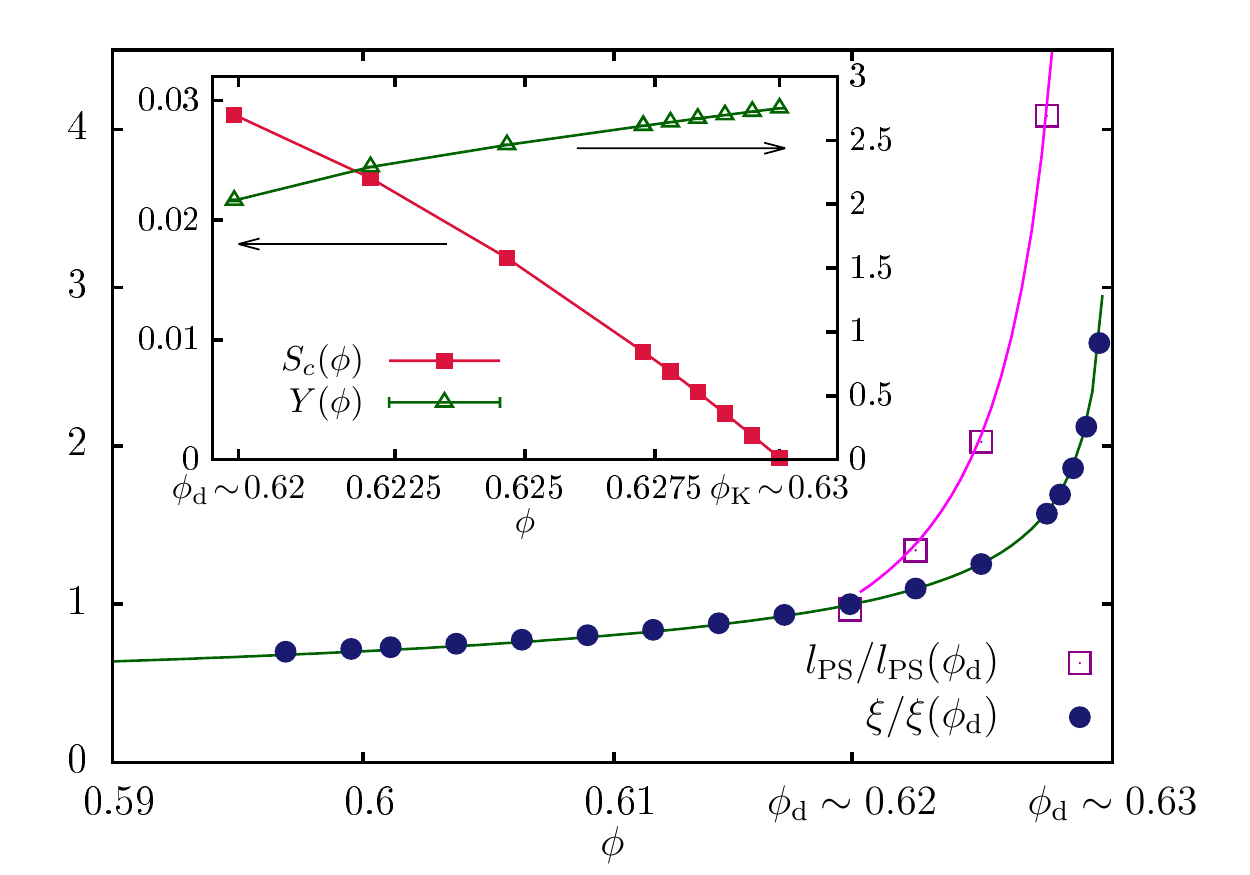}
\caption{Growth of non-trivial correlation length-scales of amorphous
  order. Small corrections to the power law behavior with exponent $-1/3$ are
  present in the pre-asymptotic growth of $\xi$. The much faster
  increase of $l_{PS}$ follows a power law divergence with exponent
  $-1$. Inset: configurational entropy and an indirect result on the amorphous surface tension of the unpinned system. \label{corrlength}} 
  \vspace{-0.6cm}
\end{figure}

According to the Random First Order Transition (RFOT) theory, above a dynamical crossover density $\phi_d$ 
(or below $T_d$), thermodynamics of dense granular systems (or 
deeply supercooled liquids) starts to be dominated by a large number of 
particularly stable amorphous configurations
corresponding to specific local rearrangements. 
A non-zero entropy of the $\mathcal{N}$ stable amorphous configurations can be 
defined as $S_c=\lim_{N\rightarrow\infty}\log(\mathcal{N})/N\neq0$ and is called 
configurational entropy.
Further increasing the density (or supercooling) the entropy of stable 
amorphous configurations decreases and vanishes at finite concentration 
(or temperature) leading in the RFOT theory to the occurrence at $\phi_K$ (or
$T_K$) of the IGT, a singularity of the thermodynamic entropy (or free-energy)
potential.\\ 
The classical theory of liquids based on diagrammatic expansion of the 
Morita-Hiroike (M-H) potential\cite{morhir61,hanmcd90}
has been adapted
to capture the effects of this multi-state scenario and the occurrence of the
IGT. 
Among other approximation schemes adopted to compute pair
correlation functions, HNC corresponds to a
stationary point of a truncated M-H potential where two-line
irreducible diagrams have been neglected\cite{hanmcd90}.\\ 
A fundamental step in the formulation of a HNC theory of glass-formers is the
introduction of a number of copies (or replicas) of the system and the study of pair
correlation functions between particles of different replicas.
In practice a system is considered composed by a mixture of particles from 
different $m$ copies of the original glass-former with positions given by the $mN$
vectors in $3$ dimensions $\{{x}_i^a\}$, 
where $i\in[1,N]$ is the particle index and $a\in[1,m]$ the replica index. 
Particles of the same replica interact through the usual pair-potential
$\psi(r)$ of the chosen model glass-former, while particles from different
replicas do not see each other.\\
For $m=1$ the problem becomes a standard HNC liquid
computation in terms of usual pair correlation function
$g(x,y)=V^2/N^2\sum_{i\neq j}\langle \delta(x-x_i^a) \delta(y-y_j^a)\rangle$.
A pair correlation function between particles of different replicas also appears in the general case:
$\widetilde{g}(x,y)=V^2/N^2\sum_{i, j}\langle \delta(x-x_i^a)\delta(y-y_j^b)\rangle$
with $a\neq b$. 
Thermodynamic potentials of the glass-former as a function of $\phi$ (or $T$)
can be computed using the M-H potential of replicated system $S^{\text{M-H}}_m$. 
The entropy of glass-former\cite{monass95} is 
$s = \left.\partial S^{\text{M-H}}_m/\partial m\right|_{m=1}$ and the
configurational entropy\cite{monass95} is
$S_c=\left.-m^2\partial[m^{-1}S^{\text{M-H}}_m]/\partial m\right|_{m=1}$.
\\
In the liquid phase, pair correlation functions among different replicas are
trivially equal to one, indicating that replicas are always completely independent.
As soon as thermodynamics starts to be dominated by particularly
stable configurations, a metastable solution
(a new stationary point of the M-H truncated potential) with a non trivial
$\widetilde{g}(x,y)$ structure appears. 
Still, particles of different replicas do not directly interact, but the glassy multi-state
structure forces different copies of
the system to lie in the same stable configuration and originates effective inter-replica couplings.\\ 
To study RPGT, we deal with a Hard Sphere system where a fraction $c$
of particles are frozen in an equilibrium reference configuration.
We then study the thermodynamics of the remaining free particles
replicated $m$ times to probe the formation of a glassy multi-state
structure. Finally we average over the equilibrium configurations of
frozen particles. In practice, to realize this construction without breaking
translational invariance,
we will consider a mixture of $N(1-c)$ particles replicated $m$ times ($m$
species), such that particles interact only within the same specie
$\psi_{a,a}(r)=\psi(r)$, and $\psi_{a,b}(r)=0$, and an 
additional specie (the $0$th one) of $Nc$ particles that interact with
all particles in the system: $\psi_{0,a}(r)=\psi(r)$.
The $m$ copies of $N(1-c)$ particles will hence freely reorganize 
in presence of the same (pinned) equilibrium template provided by the $Nc$
non-replicated particles\footnote{We are allowed to take equilibrium averages 
of the equilibrium template and of the $m$ copies of remaining particles on
the same footing as long as we work at $\phi>\phi_K$ (or $T>T_K$)}.\\
The entropy of this mixture of $m+1$ species can be expressed as prescribed by
the M-H potential for mixtures (see
Ref.\onlinecite{morhir61,hanmcd90,parzam10}) in terms of densities
$\rho_{\alpha}$ with $\alpha\in[0,m]$, of pair correlation functions
$g_{\alpha\beta}(x,y)$ and of the Fourier Transform (FT) of
$h_{\alpha\beta}(x,y)=g_{\alpha\beta}(x,y)-1$. 
These functions are determined by the HNC equations 
\begin{equation}
\log{g_{\alpha\beta}(x,y)}+\psi_{\alpha\beta}(x,y)=h_{\alpha\beta}(x,y)-c_{\alpha\beta}(x,y)
\end{equation}
and by the Ornstein-Zwernicke closures
\begin{equation}
h_{\alpha\beta}(x,y)=c_{\alpha\beta}(x,y)+\sum_{\gamma} \int dw
h_{\alpha\gamma}(x,w)\rho_{\gamma}c_{\gamma\beta}(w,y) \ .
\end{equation}
For $m=1$, assuming symmetry among the $m$ copies of the free particles, these
sets of equations simplify and as in the unconstrained case only two different
pair correlation functions, $g(x,y)$ and $\widetilde{g}(x,y)$, appear:
\begin{eqnarray}
\log{g(x,y)}+\psi(x,y)=h(x,y)-c(x,y) 
\label{HNCg}
\\
\log{\widetilde{g}(x,y)}=\widetilde{h}(x,y)-\widetilde{c}(x,y) \ ,
\label{HNCTg}
\end{eqnarray}
with
\begin{eqnarray}
&&h(x,y)-c(x,y)=\rho\int dw h(x,w)c(w,y)
\label{OZg}
\\
&\phantom{.}&\hspace{-5mm}\widetilde{h}(x,y)-\widetilde{c}(x,y)=h(x,y)-c(x,y)+ \label{OZTg}\\
&-&(1-c)\rho\int dw [h(x,w)-\widetilde{h}(x,w)][c(w,y)-\widetilde{c}(w,y)] \ . \nonumber
\end{eqnarray}
The first ones imply a solution $g(r)$ independent from $\widetilde g(r)$ and
identical to the simple liquid solution\footnote{Pair correlations among
  particles of the same replica are not changed by the presence of other
  replicas, nor by the presence of additional common non-replicated particles,
  as long as these are also equilibrated.}.
The second ones admit two different solutions, $\widetilde g_\mathrm{L}(r)$ and
$\widetilde g_\mathrm{G}(r)$, both present in some ranges of the control
parameters $\phi$ (or $T$) and $c$.
These two solutions encode respectively the liquid (L) low
correlations between
particles of different replicas induced by the presence of a fraction of
$Nc$ particles, and the glass (G) high correlations effectively
generated by the multi-state structure.\\
In the M-H entropy expression we deal with matrices of pair correlation
functions $\sqrt{\rho_{\alpha}\rho_{\beta}}g_{\alpha\beta}$ with only four
different elements that, assuming translational and rotational
invariance, is: 
$\rho_0g_{00}=c\rho g(r)$, if $\alpha\neq0$
$\sqrt{\rho_0 \rho_{\alpha}}g_{0\alpha}=\sqrt{\rho_0\rho_{\alpha}}g_{\alpha 0}=\sqrt{c(1-c)}\rho g(r)$, 
and $\rho_{\alpha}g_{\alpha \alpha}=(1-c)\rho g(r)$, and if also $\beta\neq0$
$\sqrt{\rho_{\alpha}\rho_{\beta}}g_{\alpha\beta}=\sqrt{\rho_{\alpha}\rho_{\beta}}g_{\beta\alpha}=(1-c)\rho \widetilde{g}(r)$.
Hence the entropy of glass-former, obtained from the M-H potential of replicated system $S^{\text{M-H}}_m$, reads
\begin{widetext}
\begin{eqnarray}
s[\phi,c;g,\widetilde g]=-\frac{\rho}{2}\int \hspace{-1mm}4\pi r^2dr (1-c)\Big[(1+c)g(r)(\log(g(r))-1)+(1-c)\widetilde
  g(r)(\log(\widetilde g(r))-1)+(1+c)\psi(r)g(r)+2\Big] +\hspace{2mm}\\
\nonumber +\frac{1}{2\rho}\int_q(1-c)
\left[-(1+c)\rho  h+c\rho \widetilde h+\frac{1}{2}(1+c)\rho^2 h^2
+\frac{1}{2}(1-c)\rho^2 \widetilde h^2+\rho\frac{ch+(1-c)\widetilde
  h}{1+\rho h}+\frac{\log\Big(1+(1-c)\rho(h-\widetilde h)\Big)}{1-c}\right] \ ,
\end{eqnarray}
\end{widetext}
where $h$ and $\widetilde h$ are 
the FTs of
\begin{figure}[h]
\hspace{-0.3cm}
\includegraphics[height=0.35\textwidth, angle=0]{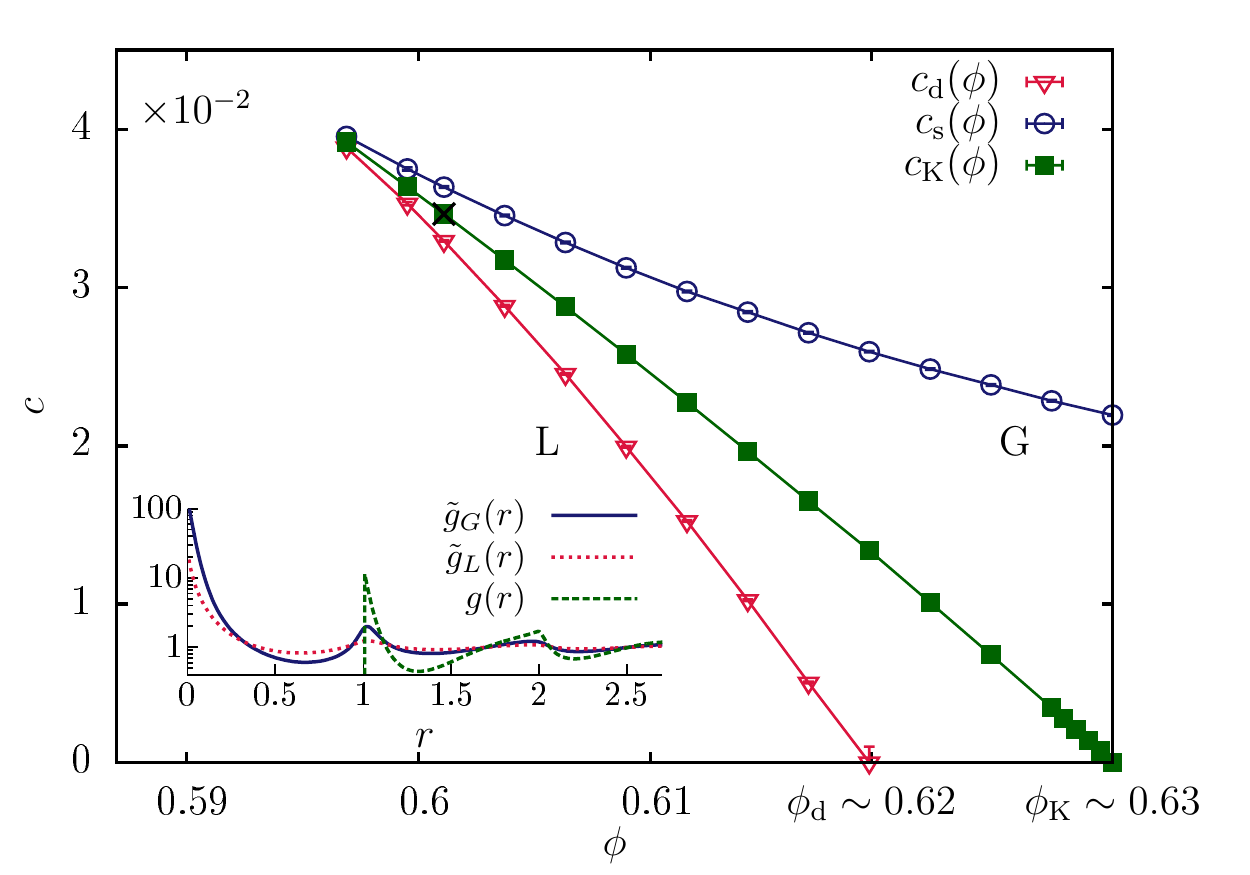}
\caption{Phase diagram of the randomly pinned Hard Sphere model in HNC
  approximations. A RPGT line is expected to begin in the IGT of unconstrained
  system and end in a critical glass transition point $\{c^*,\phi^*\}$. Two
  spinodal lines of the liquid $L$ ($c_s(\phi)$) and ideal glass $G$
  ($c_d(\phi)$) phases are also reported. The inset shows the $g(r)$, $\widetilde g_\mathrm{L}(r)$ and $\widetilde g_\mathrm{G}(r)$ in correspondence of the cross in the phase diagram ($\phi=0.601>\phi^*$ and $c_K(\phi=0.601)$). 
  \label{phasediag}} 
  \vspace{-0.2cm}
\end{figure}
$h(x)$ and $\widetilde h(x)$, and $\int_q$ represents integration in the
$3d$ momentum space\footnote{Trivial entropic terms due to single particle
  densities are neglected cuase they do not contribute to the RPGT.
  In particular, even in the $m=1$ case, exchange entropy
  is affected by the fraction of pinned particles, but this is also the case
  for experiments or simulations.}.\\
Through numerical iterative solution\footnote{We thank F. Zamponi for
sharing the C++ code for non-pinned systems.} of Eqs.(\ref{HNCg}-\ref{OZTg})
we can obtain: $\widetilde g_\mathrm{L}(r)$ using $\widetilde g(r)=1$ as initial condition, and $\widetilde g_\mathrm{G}(r)$ when the initial condition is the
non-trivial $\widetilde g(r)$ solution
of unconstrained systems.
We can also compute the entropy and hence the stability of the
corresponding phases.
For $\phi\in(\phi_d,\phi_K)$ and $c=0$ the entropy of $L$ is larger than the entropy of $G$ and the $L$
solution is stable.
When $c$ increases, for fixed $\phi$, 
the difference in entropy between the two
phases decreases, vanishes at $c_K(\phi)$, and eventually changes sign for
$c>c_K(\phi)$ where the $G$ solution becomes stable (see Fig.\ref{phasediag}).
Finally, the $L$ phase disappears beyond the spinodal line\footnote{Note that this
  spinodal do not correspond to the spinodal of threshold states $c_{d_f}$
  that can be obtained in dynamical analyses\cite{cambir13}. Here we follow the fate
  of the equilibrated liquid phase, which disappear at lower
  concentration $c_s<c_{d_f}$ in pinned fully-connected Mean Filed
  models.} $c_s(\phi)$.
When particles are pinned from equilibrium
configurations, the entropy (or more in general the free-energy) mismatch between 
the L and G phases coincides with the
configurational entropy of the constrained system (see appendices of
Ref.\onlinecite{cambir13}). 
Hence, the transition occurring at $c_K(\phi)$ 
is an actual entropy-vanishing transition, the RPGT, with similar features to the IGT\cite{cb12PNAS}.
For $\phi<\phi_d$, a second spinodal, $c_d(\phi)$, of the $G$ phase appears and the RPGT line continues in this low concentration regime,
indicating that in a HS model the glass solution can be generated by pinning
particles even if it was completely absent in the unconstrained system.
When $\phi$ decreases, the two spinodals $c_d(\phi)$ and $c_s(\phi)$ slowly approach the transition
line $c_K(\phi)$ and the two
solutions $\widetilde g_\mathrm{L}(r)$ and $\widetilde g_\mathrm{G}(r)$
computed at $c_K(\phi)$ approach each other (see inset of Fig.\ref{phasediag}). 
The three lines meet in a
critical point\cite{frparr2011,cb12PNAS,rizzot13,bicatt13} $\{c^*,\phi^*\}$, where $\widetilde g_\mathrm{L}(r)$ and $\widetilde g_\mathrm{G}(r)$ eventually merge.
For lower packing fraction, random pinning only induces trivial pair
correlation $\widetilde g_\mathrm{L}(r)$ among particles. \\
The HNC study of randomly pinned systems allows us to obtain the first
microscopic results on cooperative length-scales of non-trivial glass
fluctuations and an indirect evaluation of a key thermodynamic quantity of
RFOT: the free-energy cost of the matching between different amorphous stable configurations, {\it a.k.a.} the amorphous surface tension.
From $c_K(\phi)$ the glass phase is stable: the system is able to
spontaneously reconstruct the 
template configuration starting from the local constraints imposed by pinned
particles. This reveals in the template configuration the
presence of a subtle medium-range correlation that extends over length-scales 
smaller than the typical distance $\xi$ between pairs of pinned particles at
criticality: $\xi(\phi)\sim c_K^{-1/d}(\phi)$, where $d=3$ is the dimensionality of the system.
In quite perfect agreement with phenomenological scaling arguments, we obtain
for $\xi(\phi)$ vs $\phi_k-\phi$ a simple inverse cubic square behavior (see
Fig.\ref{corrlength}), ruled by the almost linear vanishing of $S_c(\phi)\sim \phi_K-\phi$,
except for mild deviations in the pre-asymptotic range.
Moreover, we notice that pinning a finite fraction of particles leads\cite{cb12PNAS} to a decrease of the
configurational entropy of the original system $S_c(\phi)$ rougly proportional to $c$, when $c$ is small.
Hence the configurational entropy for the pinned system is $S_c^P(\phi,c)\simeq S_c(\phi)-cY(\phi)$ and $Y(\phi)$ is a microscopic
configurational entropy loss due to the locally imposed constraint, a quantity
complementary\cite{cb12PNAS} to the amorphous surface tension: the interface free-energy cost between typical
amorphous configurations.
We can easily compute its value from a linear fit of the configurational
entropy decrease due to small pinning.
We consistently find that $Y$ is only
defined where $S_c(\phi)$ exists, hence above
$\phi_d$, and it moderately increases when the IGT is approached\cite{lubwol07,cacggv09}, as it is shown in the inset of Fig.\ref{corrlength}.
Finally, we can compute a second cooperative length scale, called point-to-set, $l_{PS}$. This length
was initially operatively defined by using an alternative
pinning procedure where all the particles are pinned except those in a cavity
of size $l$. Phenomenological arguments on that construction gave as a 
result\footnote{Actually in the original argument 
  a generic
  exponent $1/(d-\theta)$ ruled the $l_{PS}$ divergence\cite{boubir04} while in this
  paper we simply assume $\theta=2$.} 
$l_{PS}\sim Y(\phi)/S_c(\phi)\sim\xi(\phi)^d$   
and we can compute it having obtained $S_c(\phi)$ and $Y(\phi)$ from the present HNC computation.
In Fig.\ref{corrlength} the two cooperative length scales rescaled to $1$ at
$\phi_d$ are compared showing a much faster growth of $l_{PS}$, highlighting
the difference between the two lengths, and indicating the convenience of the
point-to-set procedure to reveal growth of amorphous order in real systems. 
Note that this result warns about actual important differences between apparently
analogous procedures to detect cooperative length scales\cite{bircam14}.

In this last part, we study the validity of the HNC picture described above via numerical simulations of HSs. 
\hspace{-0.5cm}
\begin{figure}[h]
\includegraphics[height=0.9\columnwidth, angle=0]{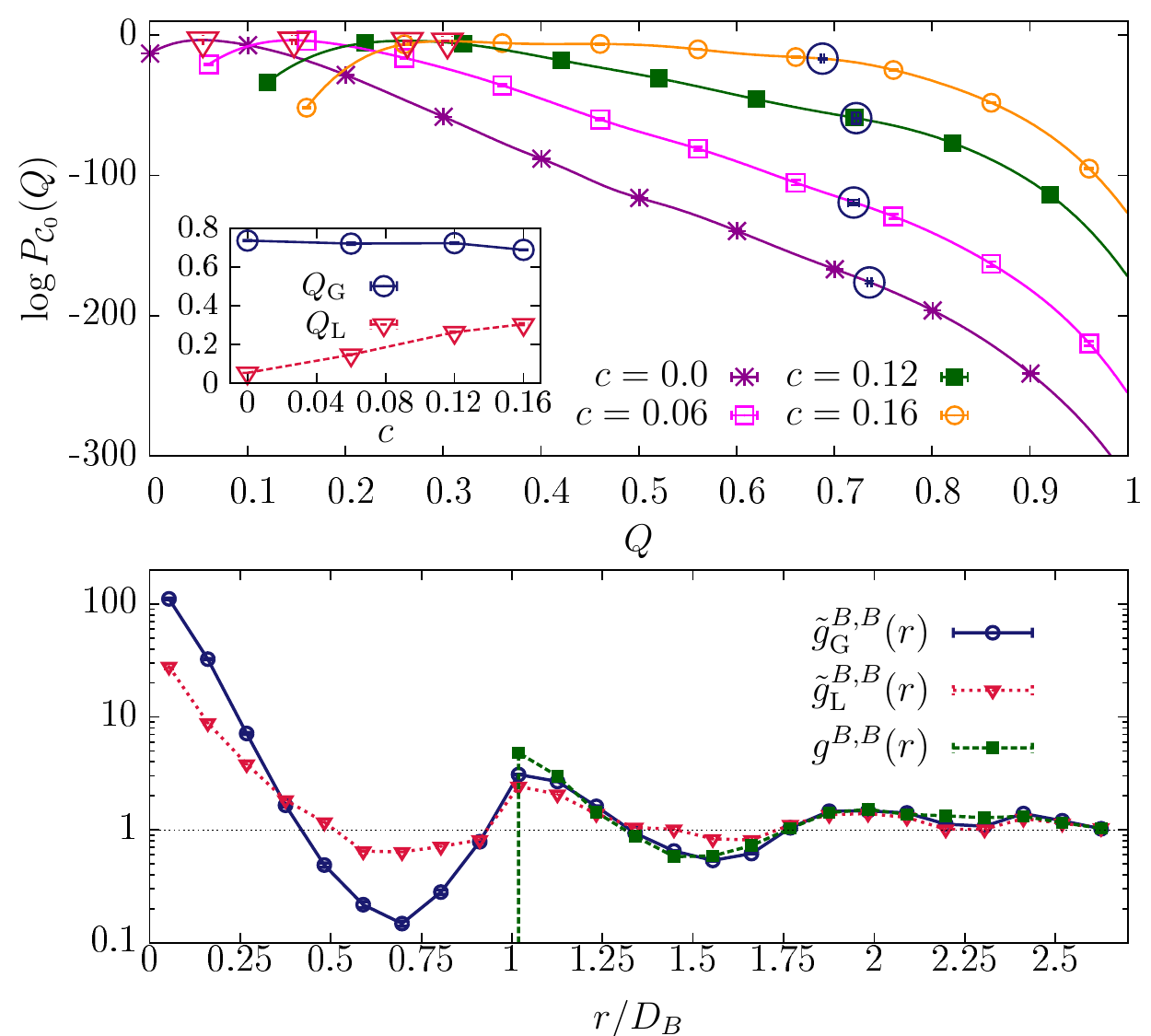}
\caption{(Top) Logarithm of $P_{\mathcal{C}_0}(Q)$ for different values of $c$. The big triangles represent $Q_\mathrm{L}$ and $Q_\mathrm{G}$ and they are plotted as function of $c$ in the inset. (Bottom) The pair correlation functions (for the $B$ particles) at the two maxima for $c=0.16$.  
  \label{gr}} 
  \vspace{-0.5cm}
\end{figure}
Despite the advantages of detecting a transition at $\phi$ well below $\phi_\mathrm{K}$, probing numerically the existence of an ideal transition line $c_\mathrm{K}(\phi)$ in equilibrium is still very hard. Indeed, dynamics at  $\phi\gtrsim\phi_\mathrm{d}$ is already very slow, and from there, characteristic times grow very fast as $c_\mathrm{K}(\phi)$ is approached. In this study we only attempt to equilibrate in the liquid part of the phase diagram in Fig.\ref{phasediag}. Still we can to obtain an indication of $c_\mathrm{K}(\phi)$ by studying the relative weight between the (metastable) glass phase $G$ and the (stable) liquid phase $L$. \\
As in the analytical calculations, we chose a reference configuration $\mathcal{C}_0$ from a well equilibrated system at $\phi=0.58$, and frozen (pin) the positions of $cN$ particles ($c=0,\ 0.06,\ 0.12$ and $0.16$). Since here we just want to show qualitatively that the random pinning does enhance the glassy behavior, we will only present results for a single reference configuration $\mathcal{C}_0$, at which we are able to thermalize up to high values of $c$ (a complete numerical study of the transition will be presented elsewhere~\cite{cammseo15}).
To tackle with thermalization issues, we apply the reversible Event Chain Monte Carlo (ECMC) algorithm~\cite{bernard09,isobe15} (with a slight variation to account for the $cN$ immobile variables~\footnote{Details on this variant of the ECMC will appear on a longer paper~\cite{cammseo15} on the simulation results.}), which represents a gain of a factor $10$ in times with respect to standard Monte Carlo (MC) moves. 
In addition to that, we used the tethered MC method~\cite{martin11} to quantify the relative weight between the L and G phases without waiting for the occurrence of spontaneous activated events during the dynamics. 
The tethered strategy~\cite{martin11} relies on independent simulations at fixed values of the order parameter, in this case the overlap $Q(\mathcal{C},\mathcal{C}_0)$ 
between the running configuration $\mathcal{C}$ and the reference one $\mathcal{C}_0$  
(the overlap is defined as $Q(\mathcal{C},\mathcal{C}_0)=\frac{1}{N}\sum_{i}^{N_\mathrm{b}}(n_{i,A}^{\mathcal{C}}n_{i,A}^{\mathcal{C}_0}+n_{i,B}^{\mathcal{C}}n_{i,B}^{\mathcal{C}_0}) $ where $n_i$ are the occupation variables of the $N_\mathrm{b}$ small boxes of size $\ell\lesssim1/\sqrt{3}D_{A,A}$: $n_i=1$ if occupied $0$ if not). One can recover the full probability distribution function, $P_{\mathcal{C}_0}(Q)=\langle\delta(Q-Q(\mathcal{C},\mathcal{C}_0))\rangle_{\mathcal{C}_0}$, being $\langle\cdot\rangle_{\mathcal{C}_0}$ the equilibrium average being $\mathcal{C}_0$ fixed, via a thermodynamic integration on the mesh of $Q$ simulation-points~\cite{parisi14}.
 We plot the logarithm of $P_{\mathcal{C}_0}(Q)$ in the top panel of Fig.\ref{gr} for several values of $c$. $P_{\mathcal{C}_0}(Q)$ always has a maximum at $Q_\mathrm{L}(c)$, and develops an elbow at high values of $Q$ that becomes more pronounced as $c$ is increased: a clear indication of the glassy enhancement upon pinning particles. Since the second glass maximum is never properly formed, as it is expected beyond mean-field, we identify the $Q_\mathrm{G}(c)$ with the position at which the high-$Q$ maximum would be if an external field $\epsilon$ coupled to the overlap were introduced to make $Q_\mathrm{L}^\epsilon(c)$ and $Q_\mathrm{G}^\epsilon(c)$ equally probable (see Refs.~\cite{parisi14,berthi13,berthier14} for technical details). The behavior of $Q_\mathrm{L}$ and $Q_\mathrm{G}$ with $c$ is shown in the inset of Fig.\ref{gr}. The IGT is given by the point at which the two maxima are equally probable, which occurs here $c_\mathrm{K}\gtrsim 0.16$. We can study the corresponding $\widetilde{g}_\mathrm{L}(r)$ and $\widetilde{g}_\mathrm{G}(r)$ at these two maxima (see bottom panel of Fig.\ref{gr}). Now $\widetilde{g}_{L,G}(r)$ is the pair-correlation function between particles in configuration $\mathcal{C}_0$ and in $\mathcal{C}$ when $Q=Q_{L,G}$. In a bidisperse system, one can compute three different $\widetilde{g}(r)$: $\widetilde{g}^{AA}_{L,G}(r)$, $\widetilde{g}^{AB}_{L,G}(r)$ and $\widetilde{g}^{BB}_{L,G}(r)$. We just show the $\widetilde{g}^{BB}_{L,G}(r)$ computed with the small particles, since $\widetilde{g}^{AA}_{L,G}(r)$ displays the same qualitative behavior. 
$\widetilde{g}^{BB}_\mathrm{L}(r)$ and $\widetilde{g}^{BB}_\mathrm{G}(r)$
show two distinct behaviours but they appear to be at the edge of merging as it is expected in proximity of the end of the RPGT line, see inset in Fig.\ref{phasediag}.

We presented the microscopic results of first principle HNC computations and of
numerical simulations in a Hard Sphere model glass-former with a fraction of
frozen particles.
The analytical and numerical results confirm expectations on the existence of a new kind of glass
transition, called RPGT, induced by pinning particles and provide
microscopic information on its occurrence in the $\phi-c$ phase diagram. 
First microscopic results on non-trivial static cooperative length scales of
glass order and on the amorphous surface tension also derive from the HNC computation. 
Finally first numerical results on $\widetilde g_{L,G}(r)$ in a bidisperse HS model
simulated with an optimized Monte-Carlo dynamics have been presented and confirm the analytic predictions on the appearance of a glass phase when pinning is increased and on the features of $\widetilde g_{G}(r)$ compared to the trivial $\widetilde g_\mathrm{L}(r)$.

\bibliographystyle{ieeetr}
\bibliography{bibliography}

\end{document}